# High precision measurement of the hyperfine splitting and ac Stark shift of the $7d\ ^2D_{3/2}$ state in atomic cesium

Bubai Rahaman and Sourav Dutta*

*Tata Institute of Fundamental Research, 1 Homi Bhabha Road, Colaba, Mumbai 400005, India*

We report the measurement of hyperfine splitting in the $7d\ ^2D_{3/2}$ state of $^{133}$Cs using high resolution Doppler-free two-photon spectroscopy in a Cs vapor cell. We determine the hyperfine coupling constants $A = 7.3509(9)$ MHz and $B = -0.041(8)$ MHz, which represent an order of magnitude improvement in the precision. We also obtain bounds on the magnitude of the nuclear magnetic octupole coupling constant $C$. Additionally, we measure the ac Stark shift of the $6s\ ^2S_{1/2} \to 7d\ ^2D_{3/2}$ transition at 767.8 nm to be $-49 \pm 5$ Hz/(W/cm$^2$), in agreement with theoretical calculations. We further report the measurement of collisional shift [$-32.6 \pm 2.0$ kHz/mTorr] and pressure broadening for the individual hyperfine levels of the $6s\ ^2S_{1/2} \to 7d\ ^2D_{3/2}$ transition. These measurements provide valuable inputs for analysis of systematic effects in optical frequency standards based on the cesium $6s\ ^2S_{1/2} \to 7d\ ^2D_{3/2}$ two-photon transition.

## I. INTRODUCTION

The cesium atom has been of interest for studies of atomic parity violation (APV) owing to its high atomic number [1–9]. A detailed understanding of APV requires high precision experiments to be supplemented with inputs from accurate theoretical calculations of the atomic structure. The simple atomic structure of Cs, containing only a single valence electron, makes it easier to perform such high precision calculations. The quality of an atomic structure calculation is often judged by its ability to produce reliable values for experimentally measurable quantities such as matrix elements, polarizabilities, Stark shifts, transition rates, hyperfine splitting (HFS) etc. In this article, we focus on precision measurement of the HFS and ac Stark shift of the $7d\ ^2D_{3/2}$ state of $^{133}$Cs. The HFS results from the electron-nucleus interaction and therefore the measurement of HFS provides insights into the electronic wave function in the vicinity of the nucleus [10,11]. Comparison of experimentally measured HFS with calculated values provides valuable insight that can then be used to further refine calculations [12,13]. The measurement of ac Stark shift enables the benchmarking of calculations of polarizabilities and is important to access the stability and systematic errors of optical frequency standards based on two-photon transitions.

In the case of $ns$ and $np$ states of cesium, the experimentally determined and theoretically calculated HFS typically agrees within 1% [13–16]. In fact, for the cesium $6s\ ^2S_{1/2}$ state, the calculations differ from the experimentally defined CODATA value only by 0.17% [15]. However, the agreement is much worse in the case of cesium $nd$ states due electron correlation effects [16,17]. On the other hand, there is interest in the cesium $nd$ state because of the work by Dzuba *et al.* [18] which pointed out that the Cs $s$-$d$ parity-nonconserving (PNC) amplitudes can be calculated with high accuracy and could compete with the accuracy of the Cs $s$-$s$ PNC amplitudes, the latter being the subject of several experimental investigations [1,2,6,8]. There also are other proposals for measurement of APV in cesium $nd$ states [19–21]. The interest in Cs $nd$ states is reflected in the several recent experimental studies pertaining in particular to the measurement of HFS [22–29]. Experimental measurements of the ac Stark shift and collisional shift of the $6s\ ^2S_{1/2} \to 7d\ ^2D_{3/2}$ transition, however, are still lacking, although they are important contributors to systematic errors. We address this gap in this article. We additionally note that a systematic comparison of experimentally measured and theoretically calculated HFS for a series of Cs $nd$ states might help in making empirical corrections to account for differences arising from unknown effects, as was proposed for $ns$ and $np$ states recently [13].

In the rest of the article, we will interchangeably use the abbreviated notation $ns_{1/2}$ for the $ns\ ^2S_{1/2}$ states, $np_J$ for the $np\ ^2P_J$ states and $nd_J$ for the $nd\ ^2D_J$ states. The hyperfine levels of the $6s_{1/2}$ and the $7d_{3/2}$ states are denoted by $F$ and $F'$, respectively.

There are several previous reports on the measurement of HFS of the Cs $7d\ ^2D_{3/2}$ state. These can be compared based on the reported value of HFS, the magnetic dipole coupling constant ($A$) and the electric quadrupole coupling constant ($B$). Belin *et al.* [22] reported the value of $A$ with a precision of 0.2 MHz using level crossing spectroscopy but the HFS was not directly measured. Kortyna *et al.* [23] used a resonant two photon excitation scheme ($6s_{1/2} \to 6p_{1/2} \to 7d_{3/2}$) in a cesium atomic beam and measured the HFS with a precision of 0.2 MHz. The nonlinearity of the laser frequency scan, calibration jitter, low signal to noise ratio (SNR) and large linewidths (~8.2 MHz) of the individual hyperfine peaks limited the accuracy of their measurement. They reported $A$ and $B$ with a precision of 0.03 MHz and 0.2 MHz, respectively. Their data did not have sufficient accuracy to determine the sign of $B$ i.e. the uncertainty was larger than the value itself. Stalnaker et al. [24] used the same excitation scheme in a cesium vapour cell but performed the experiments with a femtosecond frequency comb laser and reported the value of $A$ and $B$ with precisions of 15 kHz and 160 kHz, respectively, from which the HFS can be estimated with an average precision of ~150 kHz. Lee et al. [26] measured the HFS with a precision of ~300 kHz using Doppler-free, one color, two photon spectroscopy in a cesium vapor cell but there exists a large inconsistency between their reported HFS and their values of $A$ and $B$, as has been pointed out earlier [25]. Kumar et al. [25] used the same excitation scheme and reported the HFS with a precision of ~100 kHz. However, there are inconsistencies in their reported values, e.g. adding the $F' = 2 \leftrightarrow F' = 3$, $F' = 3 \leftrightarrow F' = 4$ and $F' = 4 \leftrightarrow F' = 5$ HFS gives 89.01 MHz which differs from the reported value of the $F' = 2 \leftrightarrow F' = 5$ HFS (88.59 MHz) by 420 kHz i.e. by four times their quoted uncertainty. It is therefore likely that their reported values of $A$ and $B$ have much higher uncertainties than those quoted, 0.01 MHz and 0.1 MHz, respectively. Recently, Wang et al. [27] reported a measurement of HFS with an accuracy of around 250 kHz, which is consistent with earlier reports but of lower precision compared to Refs. [23–25]. A comparative study of these results is provided in the supplementary files [30]. In summary, the HFS of Cs $7d\ ^2D_{3/2}$ state is known, at best, with a precision of around 150 kHz [24] and, while the reported values of $A$ agree with one another, there are large uncertainties in the reported values of $B$. One of the primary contributors to the uncertainties and inconsistencies in Refs. [23,25–27] is the nonlinearity of the laser frequency scans. Notably, there are no prior reports on the measurement of ac Stark shift and collisional shift for the cesium $6s\ ^2S_{1/2} \rightarrow 7d\ ^2D_{3/2}$ two-photon transition.

---

*sourav.dutta@tifr.res.in

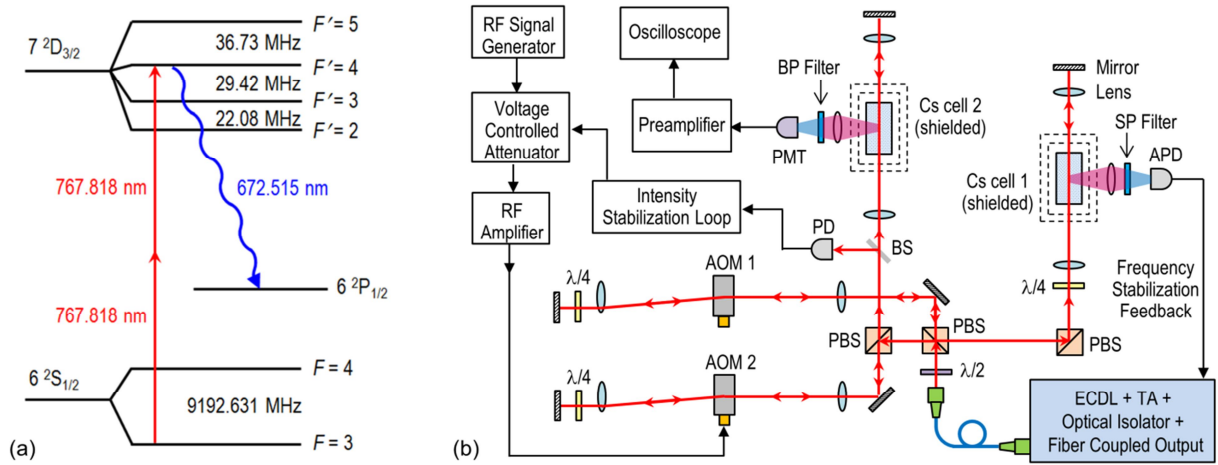

FIG. 1. (a) The energy level diagram (not to scale) of $^{133}$Cs depicting the energy levels and hyperfine splittings relevant to the experiments. The two-photon excitation and the fluorescence detection schemes are also shown. (b) A schematic diagram of the experimental setup.

In this article, we report the measurement of the HFS in the $7d\ ^2D_{3/2}$ state of atomic cesium with an average precision of around 8 kHz using Doppler-free two-photon spectroscopy. Our results provide more than an order of magnitude improvement over earlier HFS measurements. This improvement is achieved by judicious choice of the detection wavelength [see Fig. 1(a)] that enables measurement with high SNR and the implementation of a technique for precisely linear frequency scans using an acousto-optic modulator (AOM). We derive the values of $A$ and $B$ with unprecedented accuracy and estimate the value of the magnetic octupole coupling constant ($C$). Furthermore, we report the measurement of the ac Stark shift in the line position as a function of the laser power and the collisional shift as a function of cesium vapour pressure, which enable us to rule out systematic effects that often dominate the error budget in high precision measurements.

The article is organized as follows. We first discuss the experimental details and the measurement technique. We then present the experimental results, analysis of the data, extraction of the coupling constants and a comparison of our results with previous reports. We then summarize with a few concluding remarks.

## II. EXPERIMENTAL SETUP

The energy level diagram relevant to the experiment is shown in Fig. 1(a). We excite the Cs $6s\ ^2S_{1/2} \rightarrow 7d\ ^2D_{3/2}$ two-photon transition at 767.8 nm using a cw narrow-band laser system. The atoms in the $7d_{3/2}$ state de-excite following various decay routes. We record the emission at 672.5 nm arising from the $7d_{3/2} \rightarrow 6p_{1/2}$ decay which has the highest branching ratio of 0.65 [31] and leads to a strong fluorescence signal (the red coloured fluorescence is in fact visible to the naked eye). This detection scheme differs from earlier measurements which detected the $7d_{3/2} \rightarrow 6p_{3/2}$ fluorescence [23] or the $7p_J \rightarrow 6s_{1/2}$ fluorescence [24–27] from the cascaded $7d_{3/2} \rightarrow 7p_J \rightarrow 6s_{1/2}$ decay path, both of which have lower branching ratios. Detection of the $7p_J \rightarrow 6s_{1/2}$ decay has the additional disadvantage that the emitted light can be reabsorbed by the ground state Cs atoms already present in the Cs vapor cell. Our detection scheme based on the $7d_{3/2} \rightarrow 6p_{1/2}$ decay avoids the reabsorption issue and enables measurement with higher SNR.

A schematic diagram of the optical layout is shown in Fig. 1(b). The experiments are performed using Pyrex glass cells filled with cesium (manufactured by Precision Glassblowing). These Cs vapor cells are typically heated to ~135°C and enclosed in two layers of mu-metal shielding which can be degaussed in situ to reduce the stray magnetic field to the level of ~ 2 mG. The laser system consists of a commercial external cavity diode laser (ECDL) and a tapered amplifier (TA) in the master oscillator power amplifier (MOPA) configuration, followed by a 60 dB optical isolator and a fiber-coupled output. The laser beam has a Gaussian profile with a $1/e^2$ diameter of ~1.55 mm. The laser beam is split in to two parts – one part (beam 1) is used for frequency stabilization of the laser while the other part (beam 2) is used for the measurement of HFS. Beam 1 is send through AOM-1 in double-pass configuration which shifts the frequency of the beam by 2 × 106.5 MHz (this frequency is kept fixed for all the measurements). The beam is then focussed into a Cs vapour cell (Cs cell 1) and retro-reflected back. The fluorescence is collected from the side using a lens, passed through a 700-nm short-pass (SP) filter and detected on an avalanche photodiode (APD). This fluorescence signal is used to provide a feedback to the ECDL to keep the laser on resonance with the $6s_{1/2}$ ($F = 3$) → $7d_{3/2}$ ($F' = 4$) or the $6s_{1/2}$ ($F = 4$) → $7d_{3/2}$ ($F' = 4$) transition depending on whether beam 2 is set to record the $6s_{1/2}$ ($F = 3$) → $7d_{3/2}$ ($F' = 2, 3, 4, 5$) or the $6s_{1/2}$ ($F = 4$) → $7d_{3/2}$ ($F' = 2, 3, 4, 5$) spectra, respectively. Beam 2 is sent through the second Cs vapour cell (Cs cell 2) following a similar arrangement. However, there are some additional features to enable tuning of the frequency of beam 2 without changing its direction and power inside Cs cell 2. The double-pass through AOM-2 is carefully implemented in the cat's eye configuration [32] to ensure that the direction of the beam does not change as the frequency of AOM-2 is tuned. Additionally, a small fraction of the double-passed light is picked up using a quartz beam sampler (BS) and detected using a silicon photodiode (PD). The PD signal is fed to an intensity stabilization circuit which controls, via a voltage controlled attenuator, the power of the radio-frequency (rf) signal supplied to AOM-2. This keeps the power of beam 2 constant (within 0.2%) as the frequency of AOM-2 is tuned. The frequency of AOM-2 is tuned by scanning the frequency of the rf signal generator (Stanford Research Systems SG386) which provides a nearly perfect linear frequency scan with sub-Hz frequency resolution. The deviations from a perfectly linear scan contribute significantly less than 5 kHz to the errors in the measured HFS. We scan the rf frequency by 36 MHz centered around 108 MHz, which results in the beam 2 frequency being scanned by 72 MHz. Since of laser excites a two-photon transition, the net frequency scan is 144 MHz which is sufficient to tune across the entire $7d_{3/2}$ state hyperfine levels spanning ~88.2 MHz. We keep the scan rate low (1 Hz) to avoid any adverse effects that might arise from the bandwidth of the detection system. The beam is weakly focussed with a plano-convex lens of focal length 20 cm. We estimate the $1/e^2$ beam radius ($r$) at the focus to be 63±3 μm and Rayleigh range to be 16 mm. The beam is linearly polarized using a polarization beam splitter (PBS) of extinction ratio 1000:1 to avoid any systematic effects arising from optical pumping [In control experiments performed with circularly polarized light (and magnetic field of ~20 mG) that maximizes optical pumping, we observed shifts of around 12 kHz in the HFS compared to linearly polarized light]. The fluorescence is collected from a few-mm region near the focus using a lens system, passed through a 670-nm band-pass (BP) filter and detected on a photo-multiplier tube (PMT). The PMT current is amplified using a low noise current preamplifier and recorded on an oscilloscope. Simultaneously recorded on the oscilloscope are the PD signal and a reference voltage which is linearly proportional to frequency supplied by the rf signal generator. The data is recorded on the oscilloscope using one of the three modes: single scan, average of 4 scans and average of 16 scans.

## III. RESULTS AND ANALYSIS

The Cs atoms in the vapour cell are equally populated in the $6s_{1/2}$ ($F = 3$) and the $6s_{1/2}$ ($F = 4$) states. We record the HFS spectra of the $7d_{3/2}$ by exciting atoms from either of the two states. In Figs. 2(a) and 2(b) we show the representative spectra recorded from Cs cell 2 when the laser beam 2 frequency is scanned across the $6s_{1/2}$ ($F = 3$) → $7d_{3/2}$ ($F'$=2, 3, 4, 5) and the $6s_{1/2}$ ($F = 4$) → $7d_{3/2}$ ($F'$=2, 3, 4, 5) transitions, respectively. Superimposed on each figure is a fit of the spectrum to a combination of four independent Voigt profiles from which we determine the line centers of the individual peaks, the Lorentzian widths, the Gaussian widths and the peak heights, all of which were allowed to be unconstrained fitting parameters. We did not need to add a Doppler-background in the fit, but added an offset as a fitting parameter which turned out to be small (~0.01 V). We also show the residuals i.e. the difference between the data and the fitted function. Within ±200 kHz of the line centers, the root mean square residual is typically ~1% of the peak height. We note that fitting instead to a combination of four Lorentzian functions did not significantly alter the determination of the line center but the fit residuals at the wings of the peaks were slightly higher (specifically, the values of the fitted curve were slightly higher than the experimental data at the wings of the peaks). We therefore use the Voigt profile in our analysis. We repeat the measurements over several months collecting sets of data at different lasers powers and different temperatures ranging from 50°C to 175°C. Our analysis of HFS includes more than 2500 individual scans. Most of the data is collected at a temperature of 135°C where the signal is strong and does not suffer from systematic effects (see below).

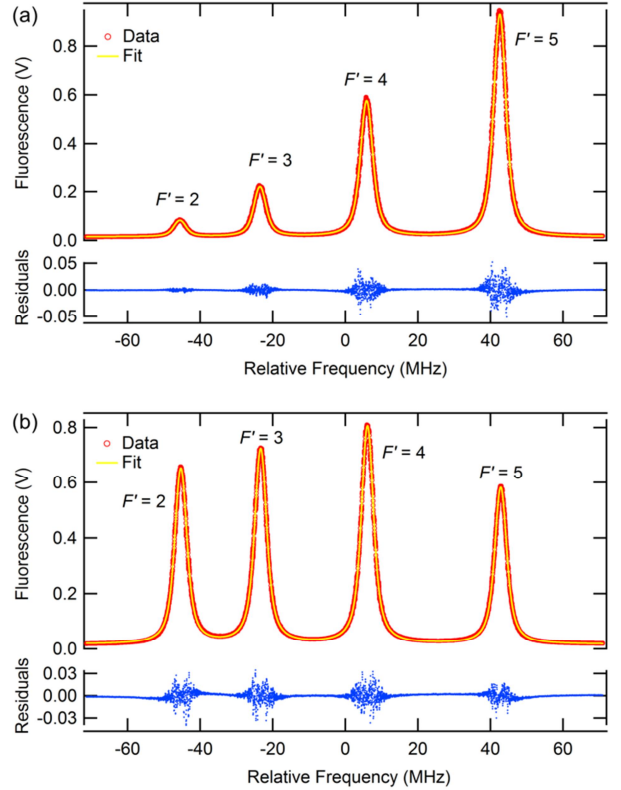

FIG. 2. Examples of Doppler-free two-photon spectra measured for (a) the $6s_{1/2}$ ($F = 3$) → $7d_{3/2}$ ($F'$=2, 3, 4, 5) transitions and (b) the $6s_{1/2}$ ($F = 4$) → $7d_{3/2}$ ($F'$=2, 3, 4, 5) transitions. The yellow line shows the fit to a combination of four independent Voigt profiles and the residuals (blue dots) show the difference between the data points and the fitted function.

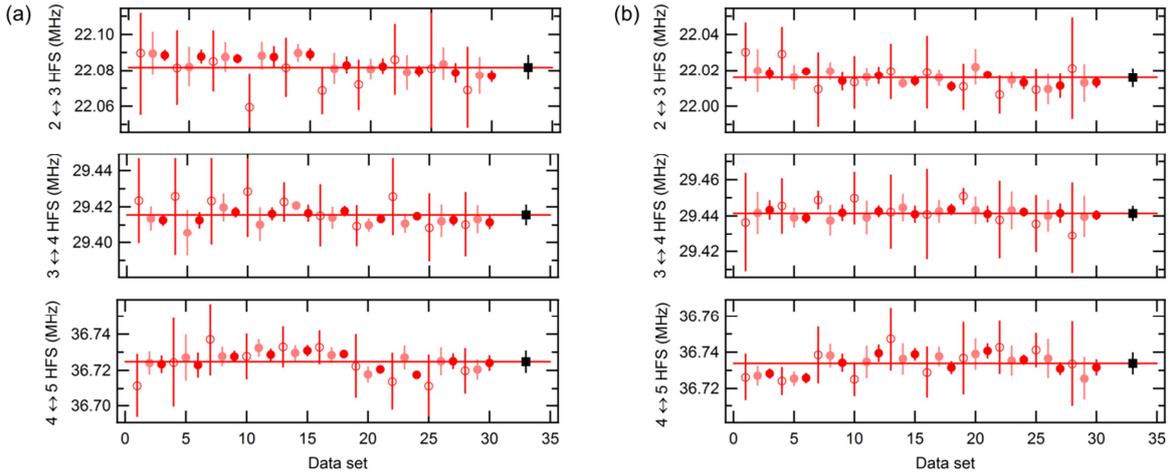

FIG. 3. The full set of data for laser power of 231 mW showing the reproducibility of the hyperfine splitting between consecutive $F'$ levels measured over several months for (a) the $6s_{1/2}$ ($F = 3$) → $7d_{3/2}$ ($F'$=2, 3, 4, 5) transitions and (b) the $6s_{1/2}$ ($F = 4$) → $7d_{3/2}$ ($F'$=2, 3, 4, 5) transitions. Each data point (circle) is an average of seven measurements and the error bars represent the standard deviation. The open circles, lightly filled circle and dark filled circles represent the data recorded using different modes of the oscilloscope viz. single scan, average of 4 scans and average of 16 scans, respectively. In each panel, the mean of the full data set is shown on the extreme right with a square symbol, with the error bar representing the standard deviation of means.

Table I: The hyperfine splitting (in MHz) measured in this work and a comparison with earlier reports.

| Hyperfine Splitting | This work $F = 3$ | This work $F = 4$ | This work[a] using $A, B, C$ | Stalnaker Ref. [24][b] | Kumar Ref. [25] | Kortyna Ref. [23] | Lee Ref. [26] | Wang Ref. [27][d] |
|---|---|---|---|---|---|---|---|---|
| $F' = 5 \leftrightarrow F' = 4$ | 36.725(6) | 36.734(6) | 36.725 | 36.80(14) | 36.93(8) | 37.0(2) | 37.28(25) | 36.85 |
| $F' = 5 \leftrightarrow F' = 3$ | 66.140(9) | 66.175(7) | 66.140 | 66.40(15) | 66.30(8) | 66.2(3)[c] | 67.49(43)[c] | 66.55[c] |
| $F' = 5 \leftrightarrow F' = 2$ | 88.222(12) | 88.191(7) | 88.222 | 88.68(19) | 88.59(11) | 88.4(3)[c] | 90.50(32)[c] | 88.88[c] |
| $F' = 4 \leftrightarrow F' = 3$ | 29.416(6) | 29.441(4) | 29.415 | 29.60(8) | 29.59(8) | 29.2(2) | 30.21(35) | 29.70 |
| $F' = 4 \leftrightarrow F' = 2$ | 51.497(9) | 51.457(6) | 51.497 | 51.88(19) | 51.79(9) | 51.4(3)[c] | 53.22(40)[c] | 52.03[c] |
| $F' = 3 \leftrightarrow F' = 2$ | 22.082(7) | 22.016(5) | 22.082 | 22.29(12) | 22.49(15) | 22.2(2) | 23.01(20) | 22.33 |

[a] These values are calculated from the values of $A$, $B$ and $C$ determined in this work (reported in the second column of Table II).
[b] These values are obtained from the values of $A$ and $B$ reported in Ref. [24].
[c] These values are deduced by adding the experimentally reported HFS between consecutive levels.
[d] The reported estimated uncertainty in these measurements is ~0.25 MHz.

In Figs. 3(a) and 3(b) we plot the measured HFS obtained from a set of spectra similar to those shown in Figs. 2(a) and 2(b), respectively. The laser power was 231 mW (measured at the input of the Cs cell 2) and Cs cell 2 temperature was 135°C for this set of data. We determine the HFS with a statistical uncertainty of ~8 kHz on the average (obtained from the scatter in the data set shown in Fig. 3) and report them in Table I along with a comparison with earlier reports.

We find a large statistically significant ($> 5\sigma$) difference of 66 kHz in the HFS between the $7d_{3/2}$ ($F' = 2$) and $7d_{3/2}$ ($F' = 3$) levels depending on whether the atoms are excited from the $6s_{1/2}$ ($F = 3$) or the $6s_{1/2}$ ($F = 4$) level. The systematic effects (ac Stark shift, collisional shift, Zeeman shift etc.) discussed below cannot account for this difference. The origin of this difference is unexplained at the moment and it will be interesting to probe this hyperfine state dependent effect further [19,20]. While determining the hyperfine constants $A$ and $B$ (discussed in the next section), we find a significant improvement in the global fit when the $F = 4 \rightarrow F' = 2$ transition is excluded from the fit, suggesting that this transition is shifted from the expected position. We will therefore put lower emphasis on this transition in our analysis; although all measured data will be presented for completeness.

Similar sets of data were collected at other values of laser powers and Cs cell 2 temperatures. We first consider the data set for different values of laser powers at a fixed value of the Cs cell 2 temperature (135°C). In Figs. 4(a) and 4(b), we plot the peak height of the individual $F'$ levels as function of the square of laser power ($P$) measured at the input of the Cs cell, along with a fit of the data to a quadratic function of $P$. The good quality of the fit over an 8× change in $P$ (i.e. 64× change in $P^2$) verifies the two-photon nature of the $6s_{1/2} \rightarrow 7d_{3/2}$ transition and attests that the data at different values laser powers are consistent with one another. In Figs. 4(c) and 4(d), we plot the line center of the individual $F'$ levels as a function of the laser power along with a fit of the data to a linear function of $P$. We find that the slopes for all the $F'$ levels in Figs. 4(c) and 4(d) are essentially the same and have an average value of $s = -1.18(4)$ MHz/W. We conclude that, while the individual $F'$ levels shift with laser power, their relative spacing and thus the HFS remain unchanged at the our level of precision (see Supplementary File [30] for plots). The systematic effect arising from the ac Stark shift contributes less than 3 kHz to the determination of HFS.

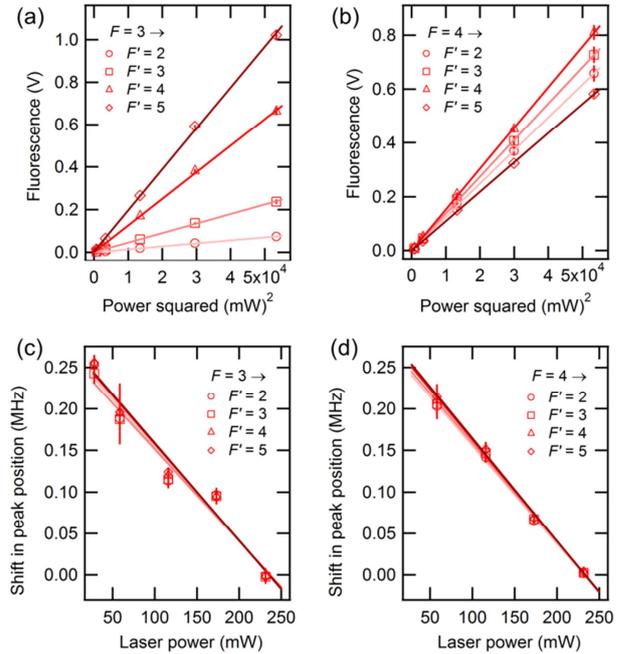

FIG. 4. (a, b) The peak fluorescence signal plotted against the square of the laser power ($P$) measured at the input of the Cs cell. The lines are fits to a quadratic function of $P$. (c, d) Shift in the peak position (relative to the data taken at 231 mW) plotted against $P$. The lines are fits to a linear function of $P$. The panels on the left and right correspond to excitation from the $6s_{1/2}$ ($F = 3$) and $6s_{1/2}$ ($F = 4$) states, respectively.

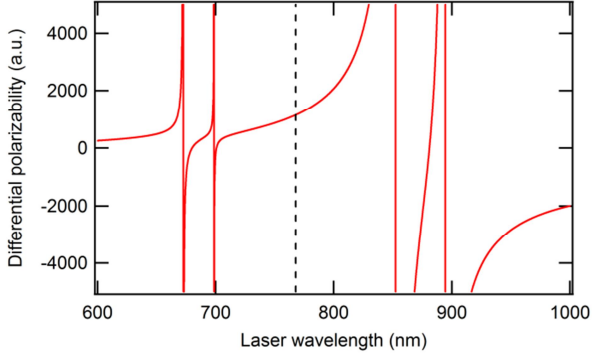

FIG. 5. The calculated differential polarizability (in atomic units, a.u.) of the $6s_{1/2} \to 7d_{3/2}$ transition plotted as a function of the laser wavelength. The dashed line indicates the wavelength (767.8 nm) at which the ac Stark shift is measured experimentally.

We also measured the HFS without the focussing lenses in front of the Cs cell 2 and the HFS were consistent with those reported in Table I. This indicates that errors from beam alignment / displacement are smaller than our measurement precision. We analyzed the Lorentzian part of the linewidths of the individual $F'$ levels at all the five laser powers and found that there is no statistically significant difference between different $F'$ levels or different laser powers i.e. power broadening is insignificant.

The measurement and characterization of the ac Stark shift is important for any optical clock based on the $6s_{1/2} \to 7d_{3/2}$ two-photon transition. We therefore determine the ac Stark shift of the $6s_{1/2} \to 7d_{3/2}$ transition per unit intensity at 767.8 nm i.e. in system-independent Hz/(W/cm$^2$) units. To do this, we calculate the peak intensity at the focus of the lens by considering the measured 86% transmission through each end of the Cs vapour cell (leading to the total bidirectional laser power of $1.5\,P$ inside the cell) and the $1/e^2$ beam radius $r = 63\pm3$ μm. Combining this with $s = -1.18(4)$ MHz/W, we determine the ac Stark shift to be $-49\pm5$ Hz/(W/cm$^2$). To compare this with theoretical results, we perform calculations of the polarizability ($\alpha$) and the ac Stark shift as outlined in Refs. [33,34], using the relevant cesium matrix elements from Ref. [31]. The matrix elements and some additional details are provided in the Supplementary File [30]. The polarizability $\alpha(6s_{1/2})$ of the $6s_{1/2}$ state at 767.8 nm is determined almost entirely by the contributions from the $6s_{1/2} \to 6p_{1/2}$ and $6s_{1/2} \to 6p_{3/2}$ transitions. In the case of the polarizability $\alpha(7d_{3/2})$ of the $7d_{3/2}$ state at 767.8 nm, the major contribution by far is from the $7d_{3/2} \to 5f_{5/2}$ transition, followed by the $7d_{3/2} \to 8p_{1/2}$ and the $7d_{3/2} \to 4f_{5/2}$ transitions. Strong cancellation of contributions from the $7d_{3/2} \to 6p_{1/2}$ and the $7d_{3/2} \to 7p_{1/2}$ transitions is noted. In Fig. 5 we plot the calculated differential polarizability $\Delta\alpha$ [$= \alpha(7d_{3/2}) - \alpha(6s_{1/2})$] of the $6s_{1/2} \to 7d_{3/2}$ transition as function of the laser wavelength. From the value of $\Delta\alpha$ (= 1152 a.u.) at 767.8 nm, we determine the ac Stark shift per unit intensity (= $-\Delta\alpha/2c\epsilon_0 h$) to be $-54$ Hz/(W/cm$^2$). This is in agreement with the experiment.

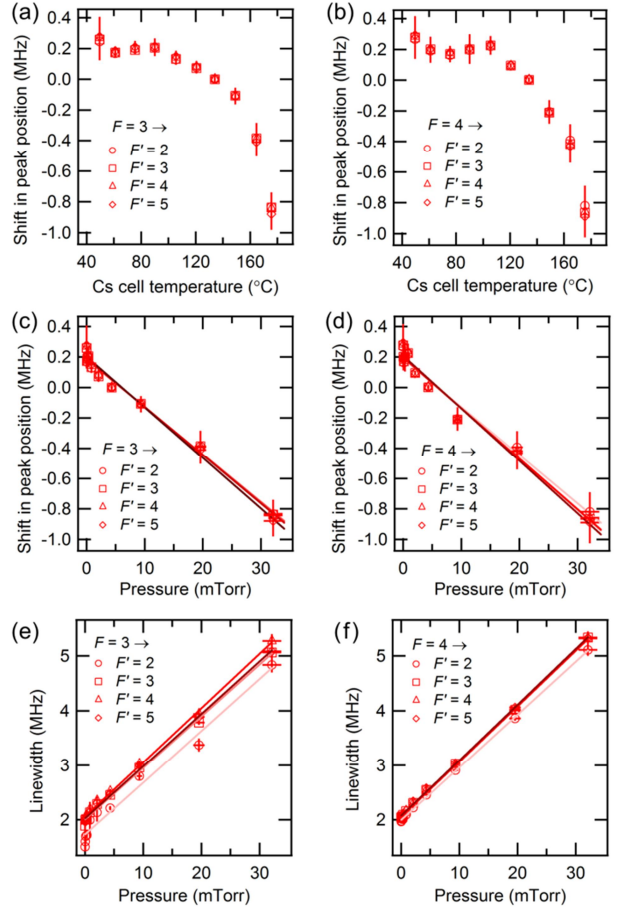

FIG. 6. (a, b) Shift in the peak position plotted against the temperature of Cs vapour cell 2. (c, d) Shift in the peak position plotted against the pressure of Cs vapour cell 2. The peak position at 135°C ($\cong 4.5$ mTorr) is taken as the reference point. (e, f) The Lorentzian part of the linewidth plotted against the pressure of Cs vapour cell 2. The lines in (c)-(f) are linear fits to the data. The panels on the left and right correspond to excitation from the $6s_{1/2}$ ($F = 3$) and $6s_{1/2}$ ($F = 4$) states, respectively.

In order to systematically study the effect of atomic collisions on the HFS, we recorded spectra at different temperatures ranging from 50°C to 175°C while keeping the laser power fixed at 231 mW. We determined the line center and linewidth from a fit to a Voigt profile as discussed above. We find that the line centers of the individual $F'$ levels shift with temperature [Figs. 6(a) and 6(b)]. However, the frequency shift is equal for all the $F'$ levels such that the HFS remains independent of the temperature. We convert the temperature ($T$) to the Cs vapour pressure $p$ using the expression [35]: $\log_{10} p = 2.881 + 4.165 - 3830/T$ and fit the frequency shift vs. $p$ data to a linear function [Figs. 6(c) and 6(d)]. We obtain the average pressure induced frequency shift to be $-32.6 \pm 2.0$ kHz/mTorr. While not of direct relevance in measurement of HFS, this pressure induced frequency shift needs to be accounted for in any optical frequency standard based on the Cs $6s_{1/2} \to 7d_{3/2}$ transition.

From the fit to Voigt profile, we extracted the Lorentzian part of the linewidths of the individual $F'$ levels and plot the full width at half maxima (FWHM) vs. $p$ in Figs. 6(e) and 6(f) along with a linear fit to the data. We find that the FWHM is the same, to within experimental uncertainty, for all the $F'$ levels. The measured linewidth of the $6s_{1/2}$ ($F = 3$) $\to 7d_{3/2}$ ($F' = 2$) transition is somewhat lower than others but this may be an artefact arising from the low SNR for this line. The FWHM increases linearly with vapour pressure with an average slope of $\sim 99(6)$ kHz/mTorr and a zero-pressure intercept of $\sim 2.04(3)$ MHz which is slightly higher than the natural linewidth ($\sim 1.7$ MHz) of the $7d_{3/2}$ state [36]. The Gaussian part of the linewidth [$\sim 2.2(1)$ MHz] does not change significantly with temperature. [A curious reader may note that on fitting the spectra to Lorentzian functions (instead of Voigt profiles) and repeating the procedure, we find a zero-pressure intercept of $\sim 3.2(1)$ MHz]. The slightly higher value of the measured linewidth [$\sim 2.04(3)$ MHz] is attributed to the combined effects of the laser linewidth (<100 kHz), transit time broadening and residual Doppler broadening arising from misalignment of the laser beams. We find that the primary source of broadening at high temperatures is collisional broadening.

## IV. HYPERFINE COUPLING CONSTANTS

At the level of first order perturbation theory, the hyperfine interaction energy of a hyperfine level $F$ is given by [10,35,37]:

$$E_F = A\frac{K}{2} + B\frac{3K(K+1) - 4I(I+1)J(J+1)}{8I(2I-1)J(2J-1)} + C\frac{5K^2(K/4+1) + K[I(I+1)+J(J+1)+3-3I(I+1)J(J+1)] - 5I(I+1)J(J+1)}{I(I-1)(2I-1)J(J-1)(2J-1)}$$

where $K = F(F+1) - I(I+1) - J(J+1)$ and $J$, $I$, $F$ are the electronic, nuclear and total angular momentum, respectively. We obtain the values of $A$, $B$ and $C$ by using the measured values of HFS as inputs to the left-hand-side of the following set of equations:

$$E_5 - E_4 = 5A + \tfrac{5}{7}B + \tfrac{4}{7}C \quad \ldots(1)$$
$$E_5 - E_3 = 9A + \tfrac{3}{7}B - \tfrac{4}{7}C \quad \ldots(2)$$
$$E_5 - E_2 = 12A - \tfrac{2}{7}B + \tfrac{4}{7}C \quad \ldots(3)$$
$$E_4 - E_3 = 4A - \tfrac{2}{7}B - \tfrac{8}{7}C \quad \ldots(4)$$
$$E_4 - E_2 = 7A - B \quad \ldots(5)$$
$$E_3 - E_2 = 3A - \tfrac{5}{7}B + \tfrac{8}{7}C \quad \ldots(6)$$

As a first approximation, we drop the terms involving $C$ and obtain the values of $A$ and $B$ from a global fit to *all* the equations and using the HFS measured from the $F = 3 \to F'$ transitions (column 2 of Table I). We obtain $A = 7.35088[51]$ MHz and $B = -0.0412[49]$ MHz. The numbers in square brackets represent the fitting errors in the last two significant digits. These values do not change appreciably if we exclude the last equation involving the 2↔3 HFS or all three equations involving the $F' = 2$ level i.e. 2↔3, 2↔4 and 2↔5 HFS, although the $F = 3 \to F' = 2$ line has the lowest SNR (see Supplementary Files [30]). The values also do not change much if the terms involving $C$ is included in the global fit. The values of $A$, $B$ and $C$ obtained by global fitting to *all* the HFS measured for excitation from the $6s_{1/2}$ ($F = 3$) are reported in the second column on Table II. These values of $A$, $B$ and $C$ reproduce each of measured HFS within the experimental uncertainty.

If we instead use the data from the $F = 4 \to F'$ transitions, and perform a global fit to *all* the equations (dropping the terms involving $C$), we get $A = 7.35107[37]$ MHz and $B = +0.0036[39]$ MHz. Further, on excluding the last equation involving the 2↔3 HFS, we get $A = 7.35123[37]$ MHz and $B = -0.0148[45]$ MHz, which are significantly different. Notice especially the change in sign of $B$. The value of $B$ changes even further to $B = -0.0628[71]$ on excluding all the three equations involving the $F' = 2$ level. On averaging the HFS measured for excitation from the $6s_{1/2}$ ($F = 3$) and $6s_{1/2}$ ($F = 4$) states and performing a global fit global fit to *all* the equations, we get $A = 7.35055[79]$ MHz and $B = -0.0311[70]$ MHz, that are more consistent with the data for excitation from the $6s_{1/2}$ ($F = 3$) state. See Supplementary Files [30] for details. Together, these suggest that the $F' = 2$ level is shifted from its expected position when atoms are excited from the $6s_{1/2}$ ($F = 4$) state. The reason for this is as yet unknown.

We therefore recommend the HFS measured for excitation from the $6s_{1/2}$ ($F = 3$) state to determine the values of $A$, $B$ and $C$. We note that the derived value of $C$ is very sensitive to the measured value of HFS. Since the fitting errors may underestimate the uncertainties, we put a stricter condition. We change one of the coupling constants (say $A$) keeping the other (say $B,C$) fixed at the reported value and note the maximum change that still reproduces all the measured HFS within their measurement uncertainties. The uncertainty obtained using this procedure is reported within parenthesis in the second column of Table II. To assess the goodness of the fits, we note that the reported value of $A$ with its uncertainty encompasses the range of values of $A$ obtained by fitting the $F = 3$ data or the average of $F = 3$ and $F = 4$ data under various constraint conditions discussed in the Supplementary File [30]. The values of $A$ and $B$ that we report represents at least an order of magnitude improvement in precision over any of the earlier reports. The value of $B$ that we report is negative and differs from zero by $\sim 5\sigma$ ($\sigma$ being the reported uncertainty). This represents a significant improvement over earlier reports, which did not differ from zero significantly. We also obtain bounds on the magnitude of $C$ that puts constraints on the magnitude of the nuclear magnetic octupole moment ($\Omega$) in $^{133}$Cs [37]. We note that including a small (of the order $\sim 1$ kHz [38]) contribution $E_F^{(2)}$ from second order perturbation theory may enable a more precise determination of $C$ and hence the value of $\Omega$ [39,40]. Such detailed calculations are beyond the scope of this work.

Table II: The hyperfine coupling constants determined in this work is reported in the second column. A comparison with earlier reports is also provided.

| Coupling constant | This work | Stalnaker Ref. [24] | Kumar Ref. [25] | Kortyna Ref. [23] | Lee Ref. [26] | Wang Ref. [27] | Theory Ref. [16] | Theory Ref. [17] |
|---|---|---|---|---|---|---|---|---|
| $A$ (MHz) | 7.3509(9) | 7.386(15) | 7.38(1) | 7.36(3) | 7.36(7) | 7.39(6) | 7.48 | 7.88 |
| $B$ (MHz) | −0.041(8) | −0.18(16) | −0.18(10) | −0.1(2) | −0.88(87) | −0.19(18) | — | — |
| $C$ (kHz) | −0.027(530) | — | — | — | — | — | — | — |

## V. CONCLUSION

We report the most precise measurement of HFS for the $7d_{3/2}$ state of Cs. Our statistical uncertainty in the HFS measurement is typically around 8 kHz. We find that the systematic errors from the ac Stark shift and the collisional shift are significantly smaller than the statistical uncertainty in our measurement. The first order Doppler-broadening is cancelled in the geometry used in the experiment and the second order Doppler shift $(v^2/c^2)f$ is small ($\sim 0.5$ kHz) and its contribution to systematic errors is negligible. The only contribution to systematic errors could come from the Zeeman effect due to the remnant $\sim 2$ mG magnetic field in the shielded Cs cell which could lead to a maximum systematic shift of $\sim 8$ kHz for circularly polarized light. In our experiments, which are performed with linearly polarized light, we found no measurable difference in the HFS even when experiments were conducted at higher magnetic fields of 15-20 mG. This implies that systematic errors in HFS arising from the Zeeman effect are also negligible at our level of precision. Adding in quadrature all the systematic errors discussed above, we find that the total systematic error is around 6 kHz which is smaller than the average statistical error in our measurements of HFS. We have therefore ignored the contribution of systematic errors in determining $A$, $B$ and $C$ reported in Table II. Nevertheless, if we add the statistical and systematic errors in quadrature for each of the measured HFS and carry out the global fitting analysis using the HFS measured from the $F = 3 \rightarrow F'$ transitions (column 2 of Table I), we obtain $A = 7.35087[64]$ MHz and $B = -0.0412[66]$ MHz. These global fitting errors are smaller than the errors reported in Table II obtained using the more stringent condition described in the previous section.

Our recommended values of $A$, $B$ and $C$ are reported in Table II. We anticipate that the values of $A$ and $B$ will not change significantly (in terms of percentage) on the inclusion of $E_F^{(2)}$ in the analysis, but the value of $C$ might change. Our work provides a motivation for the precision calculations of $E_F^{(2)}$ and $C/\Omega$ for the $7d_{3/2}$ state and combine it with HFS splitting reported in Table I to extract the value of $\Omega$. This can be used to verify the experimental results of Gerginov et al. [37] where the measured $\Omega$ was found to be in disagreement with that calculated from the nuclear shell model. Additionally, our measurement of the ac Stark shift and the collisional shift will provide valuable inputs for characterization of systematic effects in any optical frequency standard based on the cesium $6s\ ^2S_{1/2} \rightarrow 7d\ ^2D_{3/2}$ two-photon transition.

*Note added:* We refer the interested readers to our work on the Cs $7d\ ^2D_{5/2}$ state [41] and a recent review paper on hyperfine splitting of all the alkali atoms [42].

### ACKNOWLEDGMENTS
We are happy to acknowledge discussions with Bijaya K. Sahoo initiated during the ICTS Meeting on Trapped Atoms, Molecules and Ions, 2022 (code: ICTS/TAMIONs-2022/5). We acknowledge funding from the Department of Atomic Energy, Government of India under Project Identification No. RTI4002.

# Supplementary information for:
# High precision measurement of the hyperfine splitting and ac Stark shift of the 7d $^2D_{3/2}$ state in atomic cesium

**Table 1S:** Comparison of values $A$ and $B$ obtained by global fitting to the equations in the main text under different conditions e.g. using the HFS measured for excitation from $6s_{1/2}$ ($F = 3$) or the $6s_{1/2}$ ($F = 4$) state or their average. We find that the HFS measured for excitation from $6s_{1/2}$ ($F = 3$) gives a better fit and more consistent values of $A$ and $B$, which agree with the values obtained by global fitting to the average HFS. It appears that the measured line center of the $F = 4 \rightarrow F' = 2$ transition is shifted from the expected position and therefore including this line results in a worse fit for excitation from $6s_{1/2}$ ($F = 4$) state. The numbers in square brackets in this work are global fitting errors. Also tabulated are the values of HFS and coupling constants from previous reports.

| Reference | $A$ (in MHz) from global fit | $B$ (in MHz) from global fit | $F'$ levels | HFS (in MHz) determined using $A$ and $B$ | HFS (in MHz) measured and reported | Difference (in kHz) | Comments |
|---|---|---|---|---|---|---|---|
| **This work**<br><br>$A$, $B$ from $F = 3$, compared with $F = 3$ | **7.35088[51]**♦<br><br>(From $F = 3$ excitation, all HFS included in fit)* | **-0.0412[49]**♦<br><br>(From $F = 3$ excitation, all HFS included in fit)* | 5 ↔ 4<br>5 ↔ 3<br>5 ↔ 2<br>4 ↔ 3<br>4 ↔ 2<br>3 ↔ 2 | 36.725<br>66.140<br>88.222<br>29.415<br>51.497<br>22.082 | 36.725(6) [a]<br>66.140(9) [a]<br>88.222(12) [a]<br>29.416(6) [a]<br>51.497(9) [a]<br>22.082(7) [a] | 0<br>0<br>0<br>-1<br>0<br>0 | ♦Values of $A$ and $B$ reported in Table II of the main article. |
| This work<br><br>$A$, $B$ from $F = 3$, compared with $F = 4$ | 7.35088[51]<br><br>(From $F = 3$ excitation, all HFS included in fit)* | -0.0412[49]<br><br>(From $F = 3$ excitation, all HFS included in fit)* | 5 ↔ 4<br>5 ↔ 3<br>5 ↔ 2<br>4 ↔ 3<br>4 ↔ 2<br>3 ↔ 2 | 36.725<br>66.140<br>88.222<br>29.415<br>51.497<br>22.082 | 36.734(6) [b]<br>66.175(7) [b]<br>88.191(7) [b]<br>29.441(4) [b]<br>51.457(6) [b]<br>22.016(5) [b] | -9<br>-35<br>31<br>-26<br>40<br>66 | [a] Measured HFS for excitation from $6s_{1/2}$ ($F = 3$) state<br><br>[b] Measured HFS for excitation from $6s_{1/2}$ ($F = 4$) state |
| This work<br><br>$A$, $B$ from $F = 3$, compared with $F = 3$ | 7.35095[84]<br><br>(From $F = 3$ excitation, excluding $F' = 2$ in fit) | -0.0419[89]<br><br>(From $F = 3$ excitation, excluding $F' = 2$ in fit) | 5 ↔ 4<br>5 ↔ 3<br>5 ↔ 2<br>4 ↔ 3<br>4 ↔ 2<br>3 ↔ 2 | 36.725<br>66.141<br>88.223<br>29.416<br>51.499<br>22.083 | 36.725(6) [a]<br>66.140(9) [a]<br>88.222(12) [a]<br>29.416(6) [a]<br>51.497(9) [a]<br>22.082(7) [a] | 0<br>1<br>1<br>0<br>2<br>1 | * The values of $A$ and $B$ do not change on excluding the 3 ↔ 2 HFS from the global fit. |
| This work<br><br>$A$, $B$ from $F = 3$, compared with $F = 4$ | 7.35095[84]<br><br>(From $F = 3$ excitation, excluding $F' = 2$ in fit) | -0.0419[89]<br><br>(From $F = 3$ excitation, excluding $F' = 2$ in fit) | 5 ↔ 4<br>5 ↔ 3<br>5 ↔ 2<br>4 ↔ 3<br>4 ↔ 2<br>3 ↔ 2 | 36.725<br>66.141<br>88.223<br>29.416<br>51.499<br>22.083 | 36.734(6) [b]<br>66.175(7) [b]<br>88.191(7) [b]<br>29.441(4) [b]<br>51.457(6) [b]<br>22.016(5) [b] | -9<br>-34<br>32<br>-25<br>42<br>67 | |
| This work<br><br>$A$, $B$ from $F = 4$, compared with $F = 3$ | 7.35107[37]<br><br>(From $F = 4$ excitation, all HFS included in fit) | 0.0036[39]<br><br>(From $F = 4$ excitation, all HFS included in fit)♪ | 5 ↔ 4<br>5 ↔ 3<br>5 ↔ 2<br>4 ↔ 3<br>4 ↔ 2<br>3 ↔ 2 | 36.758<br>66.161<br>88.212<br>29.403<br>51.454<br>22.051 | 36.725(6) [a]<br>66.140(9) [a]<br>88.222(12) [a]<br>29.416(6) [a]<br>51.497(9) [a]<br>22.082(7) [a] | 33<br>21<br>-10<br>-13<br>-43<br>-31 | ♪ Note that the sign of $B$ has changed.<br><br>Difference between measured and calculated values are larger when $A$ and $B$ are determined using HFS data obtained by excitation from the $F = $ |
| This work<br><br>$A$, $B$ from $F = 4$, compared with $F = 4$ | 7.35107[37]<br><br>(From $F = 4$ excitation, all HFS included in fit) | 0.0036[39]<br><br>(From $F = 4$ excitation, all HFS included in fit)♪ | 5 ↔ 4<br>5 ↔ 3<br>5 ↔ 2<br>4 ↔ 3<br>4 ↔ 2<br>3 ↔ 2 | 36.758<br>66.161<br>88.212<br>29.403<br>51.454<br>22.051 | 36.734(6) [b]<br>66.175(7) [b]<br>88.191(7) [b]<br>29.441(4) [b]<br>51.457(6) [b]<br>22.016(5) [b] | 24<br>-14<br>21<br>-38<br>-3<br>35 | |
| This work<br><br>$A$, $B$ from $F = 4$, compared with $F = 3$ | 7.35123[37]<br><br>(From $F = 4$ excitation, excluding 3 ↔ 2 in fit) | -0.0148[45]<br><br>(From $F = 4$ excitation, excluding 3 ↔ 2 in fot) | 5 ↔ 4<br>5 ↔ 3<br>5 ↔ 2<br>4 ↔ 3<br>4 ↔ 2<br>3 ↔ 2 | 36.746<br>66.155<br>88.219<br>29.409<br>51.473<br>22.064 | 36.725(6) [a]<br>66.140(9) [a]<br>88.222(12) [a]<br>29.416(6) [a]<br>51.497(9) [a]<br>22.082(7) [a] | 21<br>15<br>-3<br>-7<br>-24<br>-18 | |
| This work | 7.35123[37] | -0.0148[45] | 5 ↔ 4 | 36.746 | 36.734(6) [b] | 12 | |
| $A$, $B$ from $F = 4$, compared with $F = 4$ | (From $F = 4$ excitation, excluding $3 \leftrightarrow 2$ in fit) | (From $F = 4$ excitation, excluding $3 \leftrightarrow 2$ in fit) | $5 \leftrightarrow 3$ | 66.155 | 66.175(7) [b] | -20 | 4 state. |
| | | | $5 \leftrightarrow 2$ | 88.219 | 88.191(7) [b] | 28 | |
| | | | $4 \leftrightarrow 3$ | 29.409 | 29.441(4) [b] | -32 | |
| | | | $4 \leftrightarrow 2$ | 51.473 | 51.457(6) [b] | 16 | |
| | | | $3 \leftrightarrow 2$ | 22.064 | 22.016(5) [b] | 48 | |
| This work $A$, $B$ from $F = 4$, compared with $F = 3$ | 7.35577[59] (From $F = 4$ excitation, excluding $F' = 2$ in fit) [$] | -0.0628[71] (From $F = 4$ excitation, excluding $F' = 2$ in fit) [$] | $5 \leftrightarrow 4$ | 36.734 | 36.725(6) [a] | 9 | [$]Notice the large change in values of $A$ and $B$ on excluding the $F = 4 \to F' = 2$ transition in the fit. Sign of $B$ has changed. |
| | | | $5 \leftrightarrow 3$ | 66.175 | 66.140(9) [a] | 35 | |
| | | | $5 \leftrightarrow 2$ | 88.287 | 88.222(12) [a] | 65 | |
| | | | $4 \leftrightarrow 3$ | 29.441 | 29.416(6) [a] | 25 | |
| | | | $4 \leftrightarrow 2$ | 51.553 | 51.497(9) [a] | 56 | |
| | | | $3 \leftrightarrow 2$ | 22.112 | 22.082(7) [a] | 30 | |
| This work $A$, $B$ from $F = 4$, compared with $F = 4$ | 7.35577[59] (From $F = 4$ excitation, excluding $F' = 2$ in fit) [$] | -0.0628[71] (From $F = 4$ excitation, excluding $F' = 2$ in fit) [$] | $5 \leftrightarrow 4$ | 36.734 | 36.734(6) [b] | 0 | |
| | | | $5 \leftrightarrow 3$ | 66.175 | 66.175(7) [b] | 0 | |
| | | | $5 \leftrightarrow 2$ | 88.287 | 88.191(7) [b] | 96 | |
| | | | $4 \leftrightarrow 3$ | 29.441 | 29.441(4) [b] | 0 | |
| | | | $4 \leftrightarrow 2$ | 51.553 | 51.457(6) [b] | 96 | |
| | | | $3 \leftrightarrow 2$ | 22.112 | 22.016(5) [b] | 96 | |
| This work $A$, $B$ from average of $F = 3$ and 4, compared with $F = 3$ | 7.35055[79] [*] (From average of $F = 3$ and 4, all HFS included in fit) | -0.0311[70] [*] (From average of $F = 3$ and 4, all HFS included in fit) | $5 \leftrightarrow 4$ | 36.731 | 36.725(6) [a] | 6 | [*] These values of $A$ and $B$ are consistent with those reported in Table II of the main article. |
| | | | $5 \leftrightarrow 3$ | 66.142 | 66.140(9) [a] | 2 | |
| | | | $5 \leftrightarrow 2$ | 88.215 | 88.222(12) [a] | -7 | |
| | | | $4 \leftrightarrow 3$ | 29.411 | 29.416(6) [a] | -5 | |
| | | | $4 \leftrightarrow 2$ | 51.485 | 51.497(9) [a] | -12 | |
| | | | $3 \leftrightarrow 2$ | 22.074 | 22.082(7) [a] | -8 | |
| This work $A$, $B$ from average of $F = 3$ and 4, compared with $F = 3$ | 7.35055[79] [*] (From average of $F = 3$ and 4, all HFS included in fit) | -0.0311[70] [*] (From average of $F = 3$ and 4, all HFS included in fit)) | $5 \leftrightarrow 4$ | 36.731 | 36.734(6) [b] | -3 | |
| | | | $5 \leftrightarrow 3$ | 66.142 | 66.175(7) [b] | -33 | |
| | | | $5 \leftrightarrow 2$ | 88.215 | 88.191(7) [b] | 24 | |
| | | | $4 \leftrightarrow 3$ | 29.411 | 29.441(4) [b] | -30 | |
| | | | $4 \leftrightarrow 2$ | 51.485 | 51.457(6) [b] | 28 | |
| | | | $3 \leftrightarrow 2$ | 22.074 | 22.016(5) [b] | 58 | |
| **Stalnaker** Phys. Rev. A **81**, 043840 (2010) | 7.386(15) | -0.18(16) | $5 \leftrightarrow 4$ | 36.801 | — | — | Hyperfine splittings are not reported |
| | | | $5 \leftrightarrow 3$ | 66.397 | — | — | |
| | | | $5 \leftrightarrow 2$ | 88.683 | — | — | |
| | | | $4 \leftrightarrow 3$ | 29.595 | — | — | |
| | | | $4 \leftrightarrow 2$ | 51.882 | — | — | |
| | | | $3 \leftrightarrow 2$ | 22.287 | — | — | |
| **Kumar** Phys. Rev. A **87**, 012503 (2013) | 7.38(1) | -0.18(10) | $5 \leftrightarrow 4$ | 36.771 | 36.93(8) | -159 | $5 \leftrightarrow 4 + 4 \leftrightarrow 3 + 3 \leftrightarrow 2$ gives 89.01 MHz, whereas $5 \leftrightarrow 2$ HFS is 88.59 MHz. |
| | | | $5 \leftrightarrow 3$ | 66.343 | 66.30(8) | 43 | |
| | | | $5 \leftrightarrow 2$ | 88.611 | 88.59(11) | 21 | |
| | | | $4 \leftrightarrow 3$ | 29.571 | 29.59(8) | -19 | |
| | | | $4 \leftrightarrow 2$ | 51.840 | 51.79(9) | 50 | |
| | | | $3 \leftrightarrow 2$ | 22.269 | 22.49(15) | -221 | |
| **Kortyna** Phys. Rev. A **77**, 062505 (2008) | 7.36(3) | -0.1(2) | $5 \leftrightarrow 4$ | 36.729 | 37.0 | -271 | |
| | | | $5 \leftrightarrow 3$ | 66.197 | — | — | |
| | | | $5 \leftrightarrow 2$ | 88.349 | — | — | |
| | | | $4 \leftrightarrow 3$ | 29.469 | 29.2 | 269 | |
| | | | $4 \leftrightarrow 2$ | 51.620 | — | — | |
| | | | $3 \leftrightarrow 2$ | 22.151 | 22.2 | -49 | |
| **Lee** Appl Phys B **105**, 391 (2011) | 7.36(7) | -0.88(87) | $5 \leftrightarrow 4$ | 36.171 | 37.28 | -1109 | |
| | | | $5 \leftrightarrow 3$ | 65.863 | — | — | |
| | | | $5 \leftrightarrow 2$ | 88.571 | — | — | |
| | | | $4 \leftrightarrow 3$ | 29.691 | 30.21 | -519 | |
| | | | $4 \leftrightarrow 2$ | 52.400 | — | — | |
| | | | $3 \leftrightarrow 2$ | 22.709 | 23.01 | -301 | |
| **Wang** Front. Phys. **16**, 12502 (2021) | 7.39(6) | -0.19(18) | $5 \leftrightarrow 4$ | 36.81429 | 36.85(25) | -36 | |
| | | | $5 \leftrightarrow 3$ | 66.42857 | — | — | |
| | | | $5 \leftrightarrow 2$ | 88.73429 | — | — | |
| | | | $4 \leftrightarrow 3$ | 29.61429 | 29.70(25) | -86 | |
| | | | $4 \leftrightarrow 2$ | 51.92 | — | — | |
| | | | $3 \leftrightarrow 2$ | 22.30571 | 22.30(25) | 6 | |

**Figure 1S.** The measured HFS of the $7d_{3/2}$ state at different laser powers and different days for excitation from (a) the $6s_{1/2}$ ($F = 3$) state and (b) the $6s_{1/2}$ ($F = 4$) state. The HFS do not depend on the laser power.

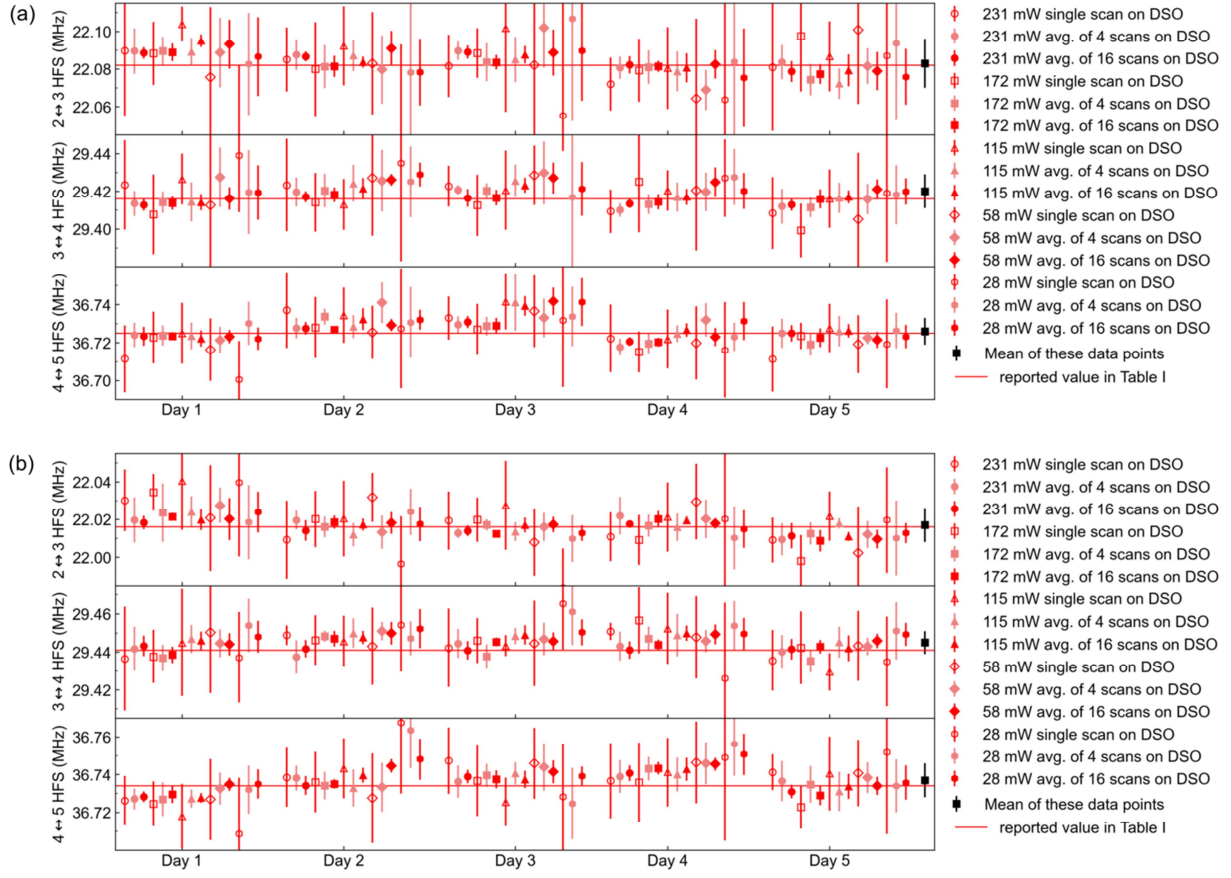

**Calculation of polarizability and ac Stark shift:**

The dynamic polarizability $\alpha(\omega, J)$ of an atom in a state with angular momentum $J$ interacting with a laser field of angular frequency $\omega$ is given by [1,2]:

$$\alpha(\omega, J) = \frac{2}{3(2J+1)} \sum_{J'} \frac{\omega_{J',J} |\langle J | d | J' \rangle|^2}{\omega_{J',J}^2 - \omega^2}$$

where $\langle J | d | J' \rangle$ is the dipole matrix element for $J \rightarrow J'$ transition, the resonant frequency for which is $\omega_{J',J}$ ($= \omega_{J'} - \omega_J$). In the polarizability of the $6s_{1/2}$ state, all $6s_{1/2} \rightarrow np_{1/2}$ and $6s_{1/2} \rightarrow np_{3/2}$ transitions contribute to the sum. The relevant matrix elements and wavelengths are listed in Table 2S [3]. In the polarizability of the $7d_{3/2}$ state, all $7d_{3/2} \rightarrow np_{1/2}$, $7d_{3/2} \rightarrow np_{3/2}$ and $7d_{3/2} \rightarrow nf_{5/2}$ transitions contribute to the sum. The relevant transition energies and matrix elements are listed in Table 2S. The differential polarizability $\Delta\alpha$ [$= \alpha(7d_{3/2}) - \alpha(6s_{1/2})$] in atomic units is calculated and plotted in Fig. 5 of the main article. The ac Stark shift per unit laser intensity ($\delta f / I$) is related to $\Delta\alpha$ through the expression:

$$\delta f / I = -\Delta\alpha / 2c\epsilon_0 h$$

where $h$ is Planck's constant, $c$ is the speed of light and $\epsilon_0$ is the permittivity of free space.

**Table 2S.** Values of transition energies and matrix elements [3]. The polarizability $\alpha(6s_{1/2})$ of the $6s_{1/2}$ state at 767.8 nm is determined almost entirely by the contributions from the $6s_{1/2} \to 6p_{1/2}$ and the $6s_{1/2} \to 6p_{3/2}$ transitions. In the case of the polarizability $\alpha(7d_{3/2})$ of the $7d_{3/2}$ state at 767.8 nm, the major contribution by far is from the $7d_{3/2} \to 5f_{5/2}$ transition, followed by the $7d_{3/2} \to 8p_{1/2}$ and the $7d_{3/2} \to 4f_{5/2}$ transitions. Strong cancellation of contributions from the $7d_{3/2} \to 6p_{1/2}$ and the $7d_{3/2} \to 7p_{1/2}$ transitions are noted.

| Transition | Energy (cm$^{-1}$) | $\langle J \| d \| J' \rangle$ (a.u.) | Contribution to $\alpha(6s_{1/2})$ at 767.8 nm (a.u.) | Transition | Energy (cm$^{-1}$) | $\langle J \| d \| J' \rangle$ in a.u. | Contribution to $\alpha(7d_{3/2})$ at 767.8 nm (a.u.) |
|---|---|---|---|---|---|---|---|
| $6s_{1/2} \to 6p_{1/2}$ | 11178.268 | 4.4978 | -370.37 | $7d_{3/2} \to 6p_{1/2}$ | -14869.566 | 2.054 | -44.57 |
| $6s_{1/2} \to 6p_{3/2}$ | 11732.307 | 6.3349 | -1077.23 | $7d_{3/2} \to 6p_{3/2}$ | -14315.527 | 0.9758 | -14.12 |
| $6s_{1/2} \to 7p_{1/2}$ | 21765.348 | 0.2781 | 0.40 | $7d_{3/2} \to 7p_{1/2}$ | -4282.490 | 6.55 | 44.42 |
| $6s_{1/2} \to 7p_{3/2}$ | 21946.395 | 0.57417 | 1.70 | $7d_{3/2} \to 7p_{3/2}$ | -4101.437 | 3.31 | 10.76 |
| $6s_{1/2} \to 8p_{1/2}$ | 25708.838 | 0.092 | 0.03 | $7d_{3/2} \to 8p_{1/2}$ | -338.999 | 31.97 | 74.77 |
| $6s_{1/2} \to 8p_{3/2}$ | 25791.509 | 0.232 | 0.20 | $7d_{3/2} \to 8p_{3/2}$ | -256.325 | 14.352 | 11.39 |
| $6s_{1/2} \to 9p_{1/2}$ | 27636.998 | 0.0429 | 0.01 | $7d_{3/2} \to 9p_{1/2}$ | 1589.162 | 9.02 | -28.30 |
| $6s_{1/2} \to 9p_{3/2}$ | 27681.676 | 0.13 | 0.06 | $7d_{3/2} \to 9p_{3/2}$ | 1633.844 | 3.564 | -4.55 |
| $6s_{1/2} \to 10p_{1/2}$ | 28726.816 | 0.0248 | 0.00 | $7d_{3/2} \to 10p_{1/2}$ | 2678.978 | 2.86 | -4.93 |
| $6s_{1/2} \to 10p_{3/2}$ | 28753.678 | 0.086 | 0.02 | $7d_{3/2} \to 10p_{3/2}$ | 2705.843 | 1.165 | -0.83 |
| $6s_{1/2} \to 11p_{1/2}$ | 29403.419 | 0.0162 | 0.00 | $7d_{3/2} \to 11p_{1/2}$ | 3355.589 | 1.554 | -1.87 |
| $6s_{1/2} \to 11p_{3/2}$ | 29420.825 | 0.0627 | 0.01 | $7d_{3/2} \to 11p_{3/2}$ | 3372.990 | 0.637 | -0.32 |
| $6s_{1/2} \to 12p_{1/2}$ | 29852.43 | 0.0115 | 0.00 | $7d_{3/2} \to 12p_{1/2}$ | 3804.597 | 1.029 | -0.95 |
| $6s_{1/2} \to 12p_{3/2}$ | 29864.341 | 0.0486 | 0.01 | $7d_{3/2} \to 12p_{3/2}$ | 3816.511 | 0.423 | -0.16 |
| $6s_{1/2} \to 13p_{1/2}$ | 30165.667 | 0.0087 | 0.00 | $7d_{3/2} \to 13p_{1/2}$ | 4117.834 | 0.755 | -0.56 |
| $6s_{1/2} \to 13p_{3/2}$ | 30174.177 | 0.0392 | 0.00 | $7d_{3/2} \to 13p_{3/2}$ | 4126.344 | 0.311 | -0.10 |
| $6s_{1/2} \to 14p_{1/2}$ | 30392.873 | 0.0069 | 0.00 | $7d_{3/2} \to 14p_{1/2}$ | 4345.038 | 0.59 | -0.37 |
| $6s_{1/2} \to 14p_{3/2}$ | 30399.165 | 0.0326 | 0.00 | $7d_{3/2} \to 14p_{3/2}$ | 4351.328 | 0.2429 | -0.06 |
| | | | | $7d_{3/2} \to 4f_{5/2}$ | -1575.607 | 13.03 | 58.55 |
| | | | | $7d_{3/2} \to 5f_{5/2}$ | 923.468 | 43.41 | -377.17 |
| | | | | $7d_{3/2} \to 6f_{5/2}$ | 2281.679 | 1.82 | -1.68 |
| | | | | $7d_{3/2} \to 7f_{5/2}$ | 3100.148 | 2.21 | -3.46 |
| | | | | $7d_{3/2} \to 8f_{5/2}$ | 3630.909 | 1.82 | -2.81 |
| | | | | $7d_{3/2} \to 9f_{5/2}$ | 3994.480 | 1.482 | -2.09 |
| | | | | $7d_{3/2} \to 10f_{5/2}$ | 4254.331 | 1.231 | -1.56 |
| | | | | $7d_{3/2} \to 11f_{5/2}$ | 4446.454 | 1.043 | -1.18 |
| | | | | $7d_{3/2} \to 12f_{5/2}$ | 4592.486 | 0.899 | -0.91 |